\documentclass[12pt,a4paper]{article}

\newcommand{\beq}{\begin{equation}}

\newcommand{\eeq}{\end{equation}}
\newcommand{\bea}{\begin{eqnarray}}
\newcommand{\eea}{\end{eqnarray}}
\newcommand{\rf}[1]{(\ref{#1})}

\newcommand{\Ee}{\mbox{E}}

\newcommand{\Kk}{\mbox{K}}

\newcommand{\om}{\omega}

\newcommand{\sn}{\mbox{sn}}
\newcommand{\dn}{\mbox{dn}}
\newcommand{\cn}{\mbox{cn}}

\begin{document}
\input{epsf}
\topmargin 0pt
\oddsidemargin 5mm
\headheight 0pt
\headsep 0pt
\topskip 9mm
\pagestyle{empty}

\begin{center}

\begin{flushright}
{\sc\footnotesize hep-th/0406176}\\
{\sc\footnotesize NORDITA-2004-46}\\
\end{flushright}
\vspace*{100pt}

\begin{center}
{\large \bf {The Circular, Elliptic Three-spin String \\
 from the $SU(3)$ Spin Chain}}
\end{center}

\vspace{26pt}

{\sl C.\ Kristjansen and T. M\aa{}nsson}
\\
\vspace{6pt}
{\sl NORDITA,} \\
{\sl Blegdamsvej 17, DK-2100 Copenhagen \O}\\

\vspace{20pt}
\end{center}

\begin{abstract}

We complete the description of the circular, 
elliptic three spin string on $AdS_5\times S^5$ having three
large angular momenta $(J_1,J_2,J_3)$ on $S^5$ 
 in the language of the integrable $SU(3)$ spin chain. First, we
recover the string solution directly from the spin 
chain sigma model and secondly, we identify the appropriate
Bethe root configuration in the so far unexplored region of parameter space.

\vfill{\noindent PACS codes: 02.30.Rz,11.15.-q,11.15.Pg,11.25.Hf,75.10.Im}

\end{abstract}

\newpage

\pagestyle{plain}

\setcounter{page}{1}
\section{Introduction}

 Semi-classical analysis of strings propagating on $AdS_5\times S^5$
 has provided a novel approach to investigating the  
AdS/CFT correspondence, the prime example being the study
of strings with several large angular momenta on $S^5$.
For such strings the classical string energy has an
 analytical dependence on the parameter $\frac{\lambda}{L^2}$ where
 $\lambda$ is the squared string tension and $L$ the total 
angular momentum.
 In addition quantum corrections to the string energy are suppressed
 as $\frac{1}{L}$ when $L\rightarrow
 \infty$~\cite{Frolov:2003qc,Frolov:2003tu}. The AdS/CFT
 correspondence~\cite{Maldacena:1997re}
 relates the energy of a IIB string state with given 
quantum numbers to the conformal dimension of a singe trace
operator of planar ${\cal N}=4$ super Yang-Mills theory with
 corresponding representation labels, mapping $\lambda$ to the 't
 Hooft coupling and $L$  to the number of constituent fields of the 
operator. This led to the suggestion that the result of the
 semi-classical string analysis should be reproduced on the gauge
 theory side by a perturbative calculation of the anomalous dimension
 followed by the limit $L\rightarrow \infty$, $\frac{\lambda}{L^2}$
 fixed --- a generalization of the BMN 
idea.
The BMN idea~\cite{Berenstein:2002jq} had triggered the development of
efficient techniques based on the use of
 effective vertices for the perturbative calculation
of anomalous dimensions of operators of 
${\cal N}=4$ SYM~\cite{Kristjansen:2002bb}. These techniques were later
 substantially improved by focusing on the dilatation generator of the
gauge theory~\cite{Beisert:2002ff,Beisert:2003tq} but their
 applicability were in practice limited to short operators  or
 operators carrying at most one large representation label such as
BMN-like operators. This limitation was overcome with the discovery
that the one loop dilatation generator of  ${\cal N}=4$ SYM could be
identified as the Hamiltonian of an integrable spin
 chain~\cite{Minahan:2002ve,Beisert:2003jj,Beisert:2003yb}. 
A connection between gauge theories and spin chains
was observed earlier in the context of QCD~\cite{Lipatov:1993yb}
and recently further integrable structures in QCD were
 revealed~\cite{Ferretti:2004ba}. In the spin chain formulation
 considering large  representation labels translates into going to the 
thermodynamical limit. 
When the number of large representation labels exceeds
one the spin chain Bethe equations~\cite{Bethe:1931hc}  turn into a set
 of integral equations involving a number  of  continuum Bethe root
 densities. In certain cases corresponding to certain sub-sectors of
${\cal N}=4$ SYM it has been possible to solve 
these equations exactly. The simplest possible closed sub-sector of
${\cal N}=4$ SYM is the $SU(2)$ sub-sector consisting of operators
composed of two out of the three complex scalar fields.
 In the $SU(2)$
sub-sector at one loop level, assuming both of the possible
 representation
labels to be large, two types of solutions of the Bethe equations were 
found and
 these were identified as the gauge theory duals of respectively a
 folded and a circular string in $AdS_5\times S^5$ having two large
 angular momenta on $S^5$~\cite{Beisert:2003xu,Frolov:2003xy}. 
The
$SU(2)$ sector remains closed to all loop orders~\cite{Beisert:2003tq}
and an extension of the spin chain picture including an appropriate
Bethe ansatz was proposed in~\cite{Serban:2004jf} to three loops, 
see also~\cite{Beisert:2004hm}. Furthermore, at one and two-loop order
 there exists
a general proof of the equivalence between solutions of the Bethe
 equations in the thermodynamical limit and solutions of the string
sigma model for large conserved 
charges~\cite{Kazakov:2004qf}. Equivalence between semi-classical
 strings and long operators has also been proved at the level of
 actions at one as well as at two loop order 
by matching continuum sigma models derived from respectively
the spin chain and the string 
theory ~\cite{Kruczenski:2003gt,Kruczenski:2004kw}.

 The study of the relation between gauge theory operators and
 semi-classical strings is less
 developed in other sub-sectors of ${\cal N}=4$ SYM. The $SU(3)$
sub-sector, consisting of operators built from the three
 complex scalars of ${\cal N}=4$ SYM is a natural place to start
 extending the analysis. At one-loop order the dilatation operator
restricted to this sub-sector is identical to the
 Hamiltonian
of an integrable $SU(3)$ spin chain, the length $L$ of the spin chain
being given by the number of constituent fields of the operators
considered.
The $SU(3)$ sub-sector is, however,
only a closed sub-sector at this order. Beyond one loop one has to
 consider the larger $SU(2|3)$ sub-sector in order to have a strictly
closed set of operators~\cite{Beisert:2003jj,Beisert:2003ys}. 
Recently, arguments
 were given, though, that the $SU(3)$ sector can be considered as
 closed
in the thermodynamical limit~\cite{Minahan:2004ds}. Generic operators
in the $SU(3)$ sub-sector are expected to be dual to strings carrying
 three
non-vanishing angular momenta $(J_1,J_2,J_3)$
on $S^5$. The first classical solution
 of the string sigma model describing such a three-spin situation was
 provided by Frolov and Tseytlin and had two out of the three spins
identical,
 i.e. $(J_1,J_2,J_3)=(J,J',J')$~\cite{Frolov:2003qc,Frolov:2003tu}.
The corresponding Bethe root configuration of the $SU(3)$ spin chain
was identified in~\cite{Engquist:2003rn}. Also fluctuations around the
classical solution has been understood from the spin chain
 perspective~\cite{Freyhult:2004iq}. Later numerous other three-spin
string solutions were found and
 classified~\cite{Arutyunov:2003uj,Arutyunov:2003za}. Briefly stated,
 three
spin string solutions can be classified as being either
 rational~\cite{Arutyunov:2003za}, elliptic or
 hyper-elliptic~\cite{Arutyunov:2003uj}. The case
 $(J_1,J_2,J_3)=(J,J',J')$ can be reached as a limiting case of the
 rational
as well as of the elliptic situation. In
 reference~\cite{Kristjansen:2004ei} the Bethe root configuration
 corresponding
to an elliptic three spin string  of circular type was identified in
 the region of parameter space where $J_2\approx J_3$, $J_1>J_2,J_3$.
In the present paper we identify the Bethe root configuration in the
 opposite limit, i.e. $J_1 \approx J_2$, $J_3<J_1,J_2$. Furthermore, we
 show how to recover the circular, elliptic three spin string directly
from the continuum $SU(3)$ spin chain sigma model, derived 
in~\cite{Hernandez:2004uw,Stefanski:2004cw}.

\section{The continuum $SU(3)$ spin chain sigma model}

Imposing the thermodynamical limit $L\rightarrow \infty$ 
and considering long wavelength excitations, the $SU(3)$ spin chain 
can be described in terms of the
following continuum sigma model action~\cite{Hernandez:2004uw,Stefanski:2004cw}
\bea
\lefteqn{
S=\frac{L}{2\pi}\int d\sigma dt \left(\dot{\alpha}+\sin^2\theta\,\dot{\phi}
+\cos^2\theta\cos(2\psi)\,\dot{\varphi}\right)}\label{sigmamodel} \\
&&
-\frac{\lambda}{4\pi L}\int d\sigma dt \left[\theta'^2+\cos^2\theta
(\psi'^2+\sin^2(2\psi)\,\varphi'^2)+\frac1{4}\sin^2(2\theta)
(\phi'-\cos(2\psi)\,\varphi')^2\right],\nonumber
\eea
with $\sigma\in [0,2\pi]$.
Here the four variables $\theta,\psi,\phi,\varphi$ are the four angles needed
to specify a coherent $SU(3)$ spin state and 
$\alpha$ is an additional overall phase\footnote{By introducing the 
variable $\alpha$ we have effectively,
in a trivial way, extended the symmetry of the spin chain to $U(3)$.}. 
The variable $\alpha$ is
redundant as regards the dynamics of the spin chain
but is useful for establishing the connection to the string sigma model
where in particular it may play a role when it comes to 
constraints. 
The model in eqn.~\rf{sigmamodel} has the conserved angular momenta
\bea
P_{\phi}=\frac{L}{2\pi}\int d\sigma \sin^2\theta, \;\;\;\;\;
\;
P_{\varphi}=\frac{L}{2\pi}\int d\sigma \cos^2\theta\cos(2\psi),
\;\;\;\;\;
\;P_{\alpha}=\frac{L}{2\pi} \int d\sigma=L,
\label{conserved}
\eea
where we notice that $P_{\alpha}$ is simply the length of the spin
chain. 
The angular variables in eqn.~\rf{sigmamodel} are conveniently chosen so
that starting from the string
metric involving $S^5$ and the decoupled time coordinate $t$ 
\beq
ds^2=-dt^2+d\theta^2+\sin^2\theta \,d\phi_3^2+
\cos^2\theta\left(d\psi^2+\cos^2\psi\, d\phi_1^2+\sin^2\psi\, d\phi_2^2\right),
\eeq
with
\beq
\phi_1=\alpha+\varphi,\hspace{0.5cm}\phi_2=\alpha-\varphi,
\hspace{0.5cm}\phi_3=\alpha+\phi,
\eeq
the same sigma model is obtained once the 
appropriate large angular momentum limit is taken. 
One can thus make the following
identification~\cite{Hernandez:2004uw}
\beq\label{identification}
P_{\phi}=J_3,\hspace{0.7cm} P_{\varphi}=J_1-J_2.
\eeq
The Hamiltonian of the model in eqn.~\rf{sigmamodel} 
is~\cite{Hernandez:2004uw}
\beq\label{Hamiltonian}
H=\frac{\lambda}{4\pi L}\int d\sigma[\theta'^2+\cos^2\theta
(\psi'^2+\sin^2(2\psi)\,\varphi'^2)+\frac1{4}\sin^2(2\theta)
(\phi'-\cos(2\psi)\,\varphi')^2].
\eeq
In order that the solutions of the sigma model
capture the cyclicity of the trace appearing in the gauge theory operators
all variables must be periodic in $\sigma$ with period $2\pi$ and
the momentum along the $\sigma$-direction
should vanish. This momentum is given by
\beq\label{psigma}
P_{\sigma}=-\frac{L}{2\pi}
\int d\sigma \left(\sin^2\theta\, \partial_{\sigma}\phi+
\cos^2\theta \cos(2\psi)\, \partial_{\sigma}\varphi+
\partial_{\sigma}\alpha\right).
\eeq
For $\theta=\phi=0$ we recover the sigma model describing the continuum limit of
the integrable $SU(2)$ spin chain. From this sigma model, one reproduces
the two-spin folded and circular string solution when
$\dot{\psi}=0$, $\dot{\varphi}=a$ 
and $\varphi'=\alpha'=0$ where $a$ is a constant
~\cite{Kruczenski:2003gt}.
In reference~\cite{Hernandez:2004uw} it was shown how to recover the  
circular,
rational three spin string solutions of~\cite{Arutyunov:2003za}  
from the continuum $SU(3)$ spin chain sigma model.
These solutions follow from the ansatz $\theta=\theta_0$, $\psi=\psi_0$ with $\theta_0$
and $\psi_0$ constant and $\varphi'=m$, $\phi'=n$ and $\alpha'=p$ with $m$, $n$ and $p$
integer.\footnote{In~\cite{Hernandez:2004uw} the variable $\alpha$ 
was left out from the  analysis.}
 The energy as a function of the spins reads
\beq
E=\frac{\lambda}{2L} \frac{1}{L^2} \left[(2m)^2J_1J_2+(n-m)^2 J_1 J_3 
+(n+m)^2J_2 J_3\right],
\eeq
and the condition $P_{\sigma}=0$ turns into
\beq
(p+m)J_1+(p-m)J_2+(n+p)J_3=0.
\eeq
In the present paper we are interested in {\it elliptic}
three spin solutions. Such solutions
follow from the ansatz $\dot{\theta}=\dot{\psi}=0$, 
$\varphi'=\phi'=\alpha'=0$ and $\dot{\phi}=a$, $\dot{\varphi}=b$
where $a$ and $b$ are constants. With this ansatz 
the momentum along the $\sigma$ direction vanishes (cf.\
eqn.~\rf{psigma}) and
the equations of motion take the form
\beq
\label{ekv:EOM1}
b\cos^2\theta\sin(2\psi)-\frac{\lambda}{2L^2}(\cos^2\theta\,\psi')'=0,
\eeq
\beq
\label{ekv:EOM2}
\frac{\lambda}{L^2}\theta''+\sin(2\theta)(a-b+2b\sin^2\psi+
\frac{\lambda}{2L^2}\psi'^2)=0.
\eeq
One simple solution to the equations is  to have $\psi$  constant and
$b=0$.
In this case $\sin \theta=\dn v\sigma$. The solution which has our
interest can be obtained by replacing this relation by the 
more general ansatz
\bea
\theta&=&\arcsin \left({\gamma \dn (v\sigma,k)}\right), \\
\psi&=&\arcsin{\beta\frac{\cn(v\sigma,k)}
{\sqrt{1-\delta^2 \dn^2(v\sigma,k)}}},
\eea
where $\gamma$, $\beta$ and $\delta$ are constants.
We then notice  that
the equations~\rf{ekv:EOM1} and~\rf{ekv:EOM2} simplify if
$\delta=\gamma$ and
if $\beta$ and $\gamma$ are related to each other as 
$\beta^2=1-\gamma^2$. In particular,
the derivative of $\psi$ takes
a very simple form
\bea
\psi'=-v\sqrt{1-\gamma^2}\sqrt{1-\gamma^2(1-k^2)}
\frac{\dn v\sigma}{1-\gamma^2\dn^2v\sigma}.
\eea
The first equation (\ref{ekv:EOM1}) now relates $b$ 
with $k$ as
\bea
\label{ekv:b}
b=\frac{v^2\lambda}{4L^2}k^2,
\eea
and the second equation is fulfilled if
\bea
b-a=\frac{v^2\lambda}{2L^2}.
\eea
Furthermore, the requirement that the angles are invariant under a shift
$\sigma$ by $2\pi$ forces $v=2\Kk/\pi$.

Making use of the relations~\rf{conserved} and~\rf{identification} we can 
now determine
the normalized spin
$j_3= J_3/L$ 
\bea
j_3=\frac1{2\pi}\int_0^{2\pi}\gamma^2\dn^2v\sigma=\
\gamma^2\,\frac{\Ee(k)}{\Kk(k)},
\eea
where it has been used that $v=2\Kk/\pi$.
Let us furthermore define $2\epsilon=(J_1-J_2)/L$. Then 
according to eqns.~\rf{conserved} and~\rf{identification} we have
\bea
2\epsilon&=&\frac1{2\pi}\int_0^{2\pi}d\sigma\left(1-\gamma^2\dn^2v\sigma-
2\beta^2\cn^2v\sigma\right) \nonumber \\
&=&\frac{2-2\gamma^2+k^2\gamma^2}{k^2}\left[1-\frac{\Ee(k)}{\Kk(k)}\right]-
(1-\gamma^2).
\eea
Using that $\gamma^2=j_3\,\Kk/\Ee$ we get a relation between 
$\epsilon$ and $j_3$ 
\bea \label{epsilon}
\epsilon=\frac1{k^2}\left[1-\frac{\Ee(k)}{\Kk(k)}\right]-\frac1{2}
+j_3\left[\left( 1-\frac1{k^2}\right)\frac{\Kk(k)}{\Ee(k)}-\frac1{2}+
\frac1{k^2}\right].
\eea
Finally, from eqn.~\rf{Hamiltonian}
we obtain an expression for the energy as a function of 
the spins
\bea
H&=&\frac{\lambda}{4\pi L}\int_0^{2\pi}d\sigma\left(
\frac{\gamma^2v^2k^4\,\sn^2v\sigma\,\cn^2v\sigma+v^2\beta^2\,
(1-\gamma^2+\gamma^2k^2)\,\dn^2v\sigma}{1-\gamma^2\dn^2v\sigma}\right)
\nonumber \\
&=&\frac{v^2\lambda}{2 L}\left[
\frac{\Ee(k)}{\Kk(k)}-\gamma^2(1-k^2)\right] \nonumber \\
&=&\frac{2\lambda}{\pi^2
  L}\left[\Ee(k)\Kk(k)+j_3(k^2-1)\frac{\Kk^3(k)}{\Ee(k)}
\right],
\label{energy}
\eea
where $k$ is supposed to be expressed via $j_3$ and $\epsilon$ using eqn.~\rf{epsilon}.
The relations~\rf{epsilon} and~\rf{energy} are exactly the relations defining
the circular, elliptic 
three spin
string~\cite{Arutyunov:2003uj,Arutyunov:2004,Kristjansen:2004ei}. 
For later convenience we note that solving eqn.~\rf{epsilon} for $k$
in terms of $j_3$ to leading order in $\epsilon$ and inserting the
solution
in eqn.~\rf{energy} we get
\beq\label{perturbative}
H=\frac{\lambda}{2L}\left(1-j_3+8\epsilon^2 \frac{1}{1+3j_3} +
{\cal O}(\epsilon^4)\right).
\eeq

\section{The discrete $SU(3)$ spin chain.}

At the discrete level, finding an eigenstate and an eigenvalue of the $SU(3)$
spin chain amounts to solving a set of algebraic equations for the Bethe 
roots. The Bethe roots come in two different types, 
reflecting the fact that the
Lie algebra $SU(3)$ has two simple roots.
Denoting the number of 
roots of the two types as $n_1$ and $n_2$ and the
roots themselves as $\{u_{1,j}\}_{j=1}^{n_1}$ and $\{u_{2,j}\}_{j=1}^{n_2}$ 
the Bethe equations read
\bea
\left(\frac{u_{1,j}+i/2}{u_{1,j}-i/2}\right)^L
&=&\prod_{k\neq j}^{n_1} 
\frac{u_{1,j}-u_{1,k}+i}{u_{1,j}-u_{1,k}-i}
\prod_{k=1}^{n_2} \frac{u_{1,j}-u_{2,k}-i/2}{u_{1,j}-u_{2,k}+i/2},
\label{Bethe1} \\
1&=&\prod_{k\neq j}^{n_2} 
\frac{u_{2,j}-u_{2,k}+i}{u_{2,j}-u_{2,k}-i}
\prod_{k=1}^{n_1} \frac{u_{2,j}-u_{1,k}-i/2}{u_{2,j}-u_{1,k}+i/2}.
\label{Bethe2}
\eea
We shall assume that $n_1\leq \frac{L}{2}$, $n_2\leq \frac{n_1}{2}$. 
The $SO(6)$ representation implied by this choice of Bethe roots is given
by the Dynkin labels $[n_1-2n_2,L-2n_1+n_2,n_1]$.  In terms of
the spin quantum numbers, assuming $J_1\geq J_2\geq J_3$ this
corresponds to $[J_2-J_3,J_1-J_2,J_2+J_3]$ or $J_1=L-n_1$, $J_2=n_1-n_2$,
$J_3=n_2    $. A given solution of the Bethe equations gives rise to an
eigenvalue of the spin chain Hamiltonian i.e.\ a one loop anomalous
dimension which is
\beq
\label{gammafirst}
\gamma=\frac{\lambda}{8\pi^2}\sum_{j=1}^{n_1}\frac{1}{(u_{1,j})^2+1/4}.
\eeq
The cyclicity of the trace is ensured by imposing 
the following constraint

\beq
1=\prod_{j=1}^{n_1}\left(\frac{u_{1,j}+i/2}{u_{1,j}-i/2}\right).
\label{cyclicity}
\eeq

Let us define
\beq
\alpha=\frac{n_1}{L},
\hspace{0.7cm} \beta=\frac{n_2}{L}.
\eeq
Then the spin quantum numbers are given by
$(J_1,J_2,J_3)=((1-\alpha)L,(\alpha-\beta)L,\beta L)$.
In references~\cite{Engquist:2003rn,Kristjansen:2004ei} the above 
Bethe equations were studied under the assumption that the roots 
$\{u_{2,j}\}_{j=1}^{n_2}$ were confined to an interval $[-ic,ic]$ on the
imaginary axis and the roots $\{u_{1,j}\}_{j=1}^{n_1}$ were 
living on two arches ${\cal C}_+$ and ${\cal C}_-$, each others 
mirror images with respect to zero, each symmetric around the real axis and
not intersecting the imaginary axis. For $c=0$ the
corresponding 
gauge theory operator is the dual of the folded string with
two large angular momenta on $S^5$~\cite{Beisert:2003xu} and for 
$c\rightarrow \infty$ the operator could be identified as the dual of the
circular string with three large angular momenta, $(J,J',J')$, 
$J>J'$ on $S^5$~\cite{Engquist:2003rn}. 
At an intermediate value of $c$ a critical
line $\beta=\beta_{crit}(\alpha)$ was located~\cite{Kristjansen:2004ei}
and it was proposed that above the critical line the operator was the
dual of the circular, elliptic three spin string of 
references~\cite{Arutyunov:2003uj,Arutyunov:2004}.
The proposal was supported by a perturbative calculation in the region
$\beta\approx\frac{\alpha}{2}$, i.e.\ $J_2\approx J_3$, $J_1> J_2,J_3$. Now, it
is known that the three spin string with angular momentum assignment
$(J',J',J)$ where $J<J'$ is characterized by the Bethe 
roots $\{u_{1,j}\}_{j=1}^{n_1}$ and $\{u_{2,j}\}_{j=1}^{n_2}$ being all
imaginary~\cite{Engquist:2003rn}. 
It is therefore natural to expect that something similar should
characterize the circular, elliptic three spin 
string with $J_1\approx J_2$, $J_3<J_1,J_2$, i.e.\ $1-2\alpha+\beta\approx 0$.
Below, we shall show that this is indeed the case.

\section{The imaginary root solution \label{imaginary}}

We assume that the Bethe
roots $\{u_{1,j}\}_{j=1}^{n_1}$ are 
all imaginary and distributed symmetrically around zero. 
Furthermore, in an interval of length of
${\cal O}(L)$ around zero the roots are equidistant, placed at the half-integer imaginary
numbers. This sub-set of the root configuration is denoted as the condensate. Outside 
the condensate
the roots are more distant. This distribution of the roots
$\{u_{1,j}\}_{j=1}^{n_1}$ is the one characteristic of the two 
spin circular string~\cite{Beisert:2003xu}. 
It ensures that the condition~\rf{cyclicity} 
is fulfilled (provided $n_1$
is odd and $L$ is even --- a constraint 
which should not affect quantities extracted in the
thermodynamical limit). 
The roots $\{u_{2,j}\}_{j=1}^{n_2}$ are likewise assumed to be imaginary
and symmetrically distributed around zero. They are furthermore assumed to be confined to 
the interval defined by the above mentioned condensate. The possibility of this configuration 
for the roots $\{u_{2,j}\}_{j=1}^{n_2}$ was pointed out in~\cite{Engquist:2003rn}. 
Rewriting the roots as $u_{1,k}=i\,q_{1,k}\,L$  and $u_{2,k}=i\,q_{2,k}\,L$, taking the logarithm
of the Bethe equations and imposing the limit $L\rightarrow \infty$ one is left with the
following set of integral equations~\cite{Engquist:2003rn} 
\beq
 2\int_s^t\hspace*{-0.55cm}-\hspace*{0.25cm}
dq'\,\frac{\sigma(q')}{q-q'}
+ 2 \int_s^tdq'\,\frac{\sigma(q')}{q+q'}
=\frac{2}{q}-8\log \frac{q-s}{q+s}
+ \int_{-v}^v dq'\frac{\rho(q')}{q-q'},\hspace{0.3cm}
s<q<t,
\label{intsigma} 
\eeq
\beq
 \int_{-v}^v\hspace*{-0.65cm}-\hspace*{0.15cm}
dq' \frac{\rho(q')}{q-q'}=2 \log\frac{s+q}{s-q}+
q \int_s^t dq'\frac{\sigma(q')}{q^2-q'^2},
\hspace{0.3cm}
-v<q<v,
\label{intrho}
\eeq
where $v<s$ and where $\rho(q)$ and $\sigma(q)$ are root densities describing respectively 
the continuum
distribution of $\{q_{2,k}\}_{k=1}^{n_2}$ and the subset of $\{q_{1,k}\}_{k=1}^{n_1}$
which are positive and lie outside the condensate. The presence of the condensate, located
at $[-s,s]$, is reflected by the appearance of the logarithmic terms  in the two equations.
The densities are normalized as
\bea
\int_{-v}^v \rho(q)dq=2\beta, \label{normbeta} \\
\int_{s}^t \sigma(q)dq=\alpha-4s. \label{normalpha}
\eea
Furthermore, the anomalous dimension can be expressed as~\cite{Beisert:2003xu}
\beq
\gamma=\frac{\lambda}{8\pi^2L}\left(\frac{4}{s}-\int_s^t dq \frac{\sigma(q)}{q^2}\right).
\label{gamma}
\eeq
In order to solve the coupled integral equations~\rf{intsigma} and~\rf{intrho} 
we shall follow the strategy of~\cite{Engquist:2003rn}, i.e.\ we express
$\sigma(q)$ in terms of $\rho(q)$ by means of eqn.~\rf{intsigma} and use the resulting expression to
eliminate $\sigma(q)$ from eqn.~\rf{intrho}. First of all, let us introduce the resolvent
corresponding to the root density $\sigma(q)$
\beq \label{resolvent}
W(q)=\int_s^t dq'\,\frac{\sigma(q')}{q-q'}\equiv W_+(q)+qW_-(q), 
\eeq
with $W_{\pm}(q)=W_{\pm}(-q)$.
The resolvent is analytic in the complex plane except for a cut
along the interval $[s,t]$. We notice that $\sigma(q)$ only enters the equation~\rf{intrho} via the 
function $qW_-(q)$ and the expression for $\gamma$ via $W_-(0)$. Thus, we do not need to
determine neither $\sigma(q)$ nor $W(q)$. We recognize the integral equation~\rf{intsigma}
as the saddle point equation of the 
${\cal O}(n)$ model on a random lattice~\cite{Kostov:fy} for $n=-2$
with the terms on the right hand side playing the role of the derivative of the potential $V(q)$,
i.e.
\beq
V'(q)=\frac{2}{q}+\int_{-v}^v dq' \rho(q') \frac{1}{q-q'}-8 \log\frac{q-s}{q+s}.
\label{potential}
\eeq
Therefore, we can immediately, following~\cite{Eynard:1995nv},
 write down a contour integral expression
for $W_-(q)$
\beq
W_-(q)=\frac{1}{2}\oint_{C}\frac{d\omega}{2 \pi i}
\frac{V'(\omega)}{q^2-\om^2}
\left\{\frac{(q^2-s^2)(q^2-t^2)}{(\om^2-s^2)(\om^2-t^2)}
\right\}^{1/2},
\label{W-}
\eeq 
where $C$ is a contour which encircles the cut $[s,t]$ but not the
other singularities of the integrand and where the endpoints $s$ and
$t$ are determined by the boundary conditions
\bea
\oint_{C}\frac{d\omega}{2 \pi i}
\frac{V'(\omega)}{(\om^2-s^2)^{1/2}(\om^2-t^2)^{1/2}}&=&0, \label{bound1}\\
\oint_{C}\frac{d\omega}{2 \pi i}
\frac{V'(\omega)\,\om^2}{(\om^2-s^2)^{1/2}(\om^2-t^2)^{1/2}}&=&2\alpha-8s. \label{bound2}
\eea
Here, the latter relation is equivalent to the normalization
condition~\rf{normalpha}. Inserting the expression~\rf{potential}
into~\rf{W-}, \rf{bound1} and~\rf{bound2} we find
\bea
qW_-(q)&=&
-\frac{1}{2 qst}\left( \sqrt{(q^2-s^2)(q^2-t^2)}-st \right) \nonumber \\
&& -\frac{q}{4}\int_{-v}^v \hspace*{-0.65cm}-\hspace*{0.15cm}
d\om\,\frac{\rho(\om)}{q^2-\om^2}
\left\{\sqrt{\frac{(s^2-q^2)(t^2-q^2)}{(s^2-\om^2)(t^2-\om^2)}}-1\right\}
\nonumber \\
&&+2q \int_{-s}^s\hspace*{-0.65cm}-\hspace*{0.15cm}
 d\om\, \frac{1}{q^2-\om^2}
\left\{\sqrt{\frac{(s^2-q^2)(t^2-q^2)}{(s^2-\om^2)(t^2-\om^2)}}-1\right\},
\eea
with the boundary conditions
\bea
&& \hspace{-1.cm}\frac{2}{st}+\int_{-v}^v d\om
\frac{\rho(\om)}{\sqrt{(s^2-\om^2)(t^2-\om^2)}}
-8 \int_{-s}^s d\om
\frac{1}{\sqrt{(s^2-\om^2)(t^2-\om^2)}}=0, \\
&&\hspace{-1.3cm}-\frac{1}{2}
\int_{-v}^v d\om
\frac{\rho(\om)\om^2}{\sqrt{(s^2-\om^2)(t^2-\om^2)}}
+4 \int_{-s}^s d\om
\frac{\om^2}{\sqrt{(s^2-\om^2)(t^2-\om^2)}}=1-2\alpha-\beta.
\eea
Furthermore, we find for $\gamma$
\bea
\gamma&=&\frac{\lambda}{8\pi^2 L}\left(\frac{4}{s}+W_-(0)\right)
\nonumber \\
&
=&\frac{\lambda}{32\pi^2 L}
\left\{\frac{16}{s}+\frac{1}{s^2}+\frac{1}{t^2}
+\int_{-v}^v \hspace*{-0.65cm}-\hspace{0.25cm}d\om\,\frac{\rho(\om)}{\om^2}
\left[\frac{st}{\sqrt{(s^2-\om^2)(t^2-\om^2)}}-1\right] \right. 
\nonumber\\
&&{}\left.-8
\int_{-s}^s \hspace*{-0.65cm}-\hspace{0.25cm}d\om \,\frac{1}{\om^2}
\left[\frac{st}{\sqrt{(s^2-\om^2)(t^2-\om^2)}}-1\right] \right\}.
\label{gammafinal}
\eea
Finally, the integral equation for $\rho(q)$ takes the form
\bea
\lefteqn{\hspace*{-1.0cm}
\int_{-v}^v \hspace*{-0.65cm}-\hspace{0.25cm} dx\,
  \frac{\rho(x)}{q-x}
\left(3+\sqrt{\frac{(s^2-q^2)(t^2-q^2)}{(s^2-x^2)(t^2-x^2)}}\right)}
\label{intfinal}
\\
&=&\frac{2}{qst}\left( st-\sqrt{(s^2-q^2)(t^2-q^2)}\right)\nonumber \\
&&+8\int_{-s}^s \hspace*{-0.65cm}-\hspace{0.25cm}
dx \,\frac{1}{q^2-x^2} \sqrt{\frac{s^2-q^2}{s^2-x^2}}
\left(\sqrt{\frac{t^2-q^2}{t^2-x^2}}-1\right), \hspace{0.5cm} -v<q<v.
\nonumber
\eea

\section{Perturbative solution for $1-2\alpha+\beta\approx 0$ }

As mentioned earlier for
$1-2\alpha+\beta=0$ the gauge theory operator in question is known to be
the dual
of the circular three-spin string of
~\cite{Frolov:2003qc,Frolov:2003tu} 
which has angular momenta 
$(J',J',J)$, $J<J'$~\cite{Engquist:2003rn}.
In the following we shall  show that as we perturb
away from $1-2\alpha+\beta=0$, the operator
becomes the gauge theory dual of the circular, elliptic three-spin
string given by eqns.~\rf{epsilon} and~\rf{energy}.

Let us define
\beq
2\epsilon=1-2\alpha+\beta,
\eeq
and let us consider $\epsilon\ll \alpha,\beta$.
In terms of angular
momenta we have
$(J_1,J_2,J_3)=(\frac{1}{2}(1-\beta+2\epsilon)L,\frac{1}{2}(1-\beta-2\epsilon)L,
\beta L)$ or
\beq
\epsilon=\frac{1}{2L}(J_1-J_2)\equiv \frac{1}{2}(j_1-j_2),\hspace{0.7cm}
\beta=\frac{J_3}{L}\equiv j_3,\hspace{0.7cm} j_3< j_1,j_2.
\label{betaj}
\eeq
As pointed out in~\cite{Engquist:2003rn}, for $\epsilon=0$, the
boundary equation~\rf{bound2} is solved by setting $t=\infty$. For a
small, non-zero value of $\epsilon$ consistency of the boundary
equations requires that $t\sim \frac{1}{\epsilon}$. Expanding the two
boundary conditions to leading order in $\epsilon$ we get
\bea
&&2s^2 \pi-\frac{1}{2}\int_{-v}^v dx\, \frac{\rho(x)
  x^2}{\sqrt{s^2-x^2}}= 2\epsilon\, t, \label{bound1eps} \\
&&4\pi=\frac{1}{s}+\frac{1}{2}\int_{-v}^v dx\,
\frac{\rho(x)}{\sqrt{s^2-x^2}}-
\frac{\epsilon}{t}.\label{bound2eps}
\eea
Working at leading order in $\epsilon$,
the first of these two equations gives us $t$ as a function of
$\epsilon$ and the second tells us how $s$ (and $v$) depend on
$\epsilon$. In particular, we see that the correction to $s$ and $v$
must be ${\cal O}(\epsilon^2)$. As we shall see we do not need to know
the explicit form of these corrections. We furthermore notice that for
symmetry reasons, corrections to the integral equation~\rf{intfinal} and
to the expression for $\gamma$, i.e.\ eqn.~\rf{gammafinal} can involve only
even powers of $\epsilon$. 
Now, expanding~\rf{intfinal} for large $t$ we find that the 
corrections of 
order $\epsilon^2$ cancel out 
due to the boundary conditions and we are left with
\beq
\int_{-v}^v \hspace*{-0.65cm}-\hspace{0.25cm}
dx \, \frac{\rho(x)}{q-x}\left(3+\sqrt{\frac{s^2-q^2}{s^2-x^2}}\right)=
\frac{2}{q}\left(1-\sqrt{1-\frac{q^2}{s^2}}\right)+{\cal O}(\epsilon^4).
\label{rhoeps}
\eeq
A similar cancellation of order $\epsilon^2$ terms takes place in the expression for 
$\gamma$ and we get\footnote{ Notice that $s$ and $v$ might still get 
$\epsilon$ corrections. 
However, as already mentioned it is not necessary to know the explicit form of these 
corrections.}.
\beq
\gamma=\frac{\lambda}{32\pi^2 L}\left[ \frac{1}{s^2}+\int_{-v}^v dq\, \rho(q)\frac{1}{q^2}
\left(\frac{s}{\sqrt{s^2-q^2}}-1\right)+{\cal O}(\epsilon^4)\right].
\label{gammaeps}
\eeq
The two equations~\rf{rhoeps} and~\rf{gammaeps} thus to the given order in $\epsilon$
take the same form as for $t=\infty$ and we can proceed using a solution strategy similar to
the one employed in that case. The new element then consists in correctly taking into
account the modified boundary conditions. 
Following~\cite{Engquist:2003rn} we introduce the new variables 
\beq
q=\frac{2s\eta}{1+\eta^2},\hspace{0.7cm} x=\frac{2s\xi}{1+\xi^2},
\eeq
with $dx\rho(x)\equiv d\xi\rho(\xi)$. In these variables the integral equation~\rf{rhoeps} takes the
form 
\beq
\int_{-\nu}^\nu \hspace*{-0.65cm}-\hspace{0.25cm} d\xi \rho(\xi) \frac{1+\xi^2}{1-\xi^2}
\left(2\,\frac{1+\eta\xi}{\eta-\xi}+\frac{\eta+\xi}{1-\eta\xi}\right)=2\eta,
\label{plaquette}
\eeq
where $\nu$ is related to $v$ by
\beq
v=\frac{2s\nu}{1+\nu^2}.
\eeq
The integral equation~\rf{plaquette} is of the type characteristic of the ${\cal O}(n)$ plaquette
matrix model studied in~\cite{Chekhov:1996xy} and an explicit expression for $\rho(\xi)$ valid
for any $n$ can be written down by contour integral techniques. However, since we do not need all
the information stored in $\rho(\xi)$ and since the present case corresponds to $n=1$ which is one of
the so-called rational points of the ${\cal O}(n)$ model~\cite{Kostov:fy,Kostov:pn,Eynard:1992cn} we shall
proceed along the lines of~\cite{Engquist:2003rn}, using a method developed in~\cite{Eynard:1992cn}.
We introduce a resolvent $F(z)$ by
\beq
F(z)=\int_{-\nu}^{\nu} d\xi \rho(\xi) \frac{1+\xi^2}{1-\xi^2}\frac{1+z\xi}{z-\xi}.
\eeq
This object is analytic in the complex plane except for a cut along the interval $[-\nu,\nu]$ and it 
has the following asymptotic behaviour as $z\rightarrow \infty$
\beq
F(z) \sim \frac{p}{z}, \hspace{0.7cm} \mbox{as} \hspace{0.7cm} z\rightarrow \infty,
\eeq
with
\beq
p=\int_{-\nu}^{\nu} d\xi \,\rho(\xi)\,\frac{(1+\xi^2)^2}{1-\xi^2}.
\eeq
The constant $p$ plays a very central role since $\gamma$ can be expressed as
\beq
\gamma=\frac{\lambda}{32\pi^2 L s^2}\left(1+\frac{p}{2}\right).
\label{gammap}
\eeq
Using the definition of $F(z)$ one can now write the boundary conditions~\rf{normbeta}, 
\rf{normalpha} and~\rf{bound1} as
\bea
F(i)&=&-8\pi i s\left(1+\frac{\epsilon}{4\pi t}\right)+2i, \\
F'(i)&=& 2\beta, \\
F''(i)&=&-2i(1-\beta) +8i\,\frac{t}{s}\,\epsilon. 
\eea
Furthermore, by using analyticity arguments as in~\cite{Eynard:1992cn,Engquist:2003rn}
one can show that the function $\omega(z)$,
defined by
\beq
\omega(z)=F(z)-\frac{4z}{3}+\frac{2}{3z},
\eeq
fulfills the following cubic equation
\beq
\om^3(z)-R(z)\om(z)=S(z),
\label{cubic}
\eeq
where
\bea
R(z)&=&\frac{4}{3}\left(z+\frac{1}{z}\right)^2-64 \pi^2 s^2\left(1+\frac{\epsilon}{2\pi t}\right), \\
S(z)&=&-\frac{16}{27}\left(z+\frac{1}{z}\right)^3+\frac{4}{3}
\left(6+3p-64\pi^2 s^2\left(1+\frac{\epsilon}{2\pi t}\right)\right)\left(z+\frac{1}{z}\right).
\eea
Now, by considering the first derivative of eqn.~\rf{cubic} we get the
following expression for $p$ in terms of $s$, $t$, $\epsilon$ and $\beta$
\beq
p=32 \pi^2 s^2 (1-\beta)\left(1+\frac{\epsilon}{2\pi t}\right)-2.
\label{p}
\eeq
Furthermore, from the second derivative of eqn.~\rf{cubic} we get an expression
for $t$ as a function of $\epsilon$ and $\beta$
\beq
t=\frac{1}{16\pi \epsilon}(1-\beta)(1+3\beta)\label{teps}.
\eeq
Finally, inserting eqns.~\rf{p} and~\rf{teps} in the expression~\rf{gammap}
for $\gamma$ we see that the $s$-dependence very neatly cancels out and
we are left with 
\beq\label{gammapert}
\gamma=\frac{\lambda }{2L} \left(1-j_3+8\epsilon^2
\frac{1}{1+3j_3}+{\cal O}(\epsilon^4) \right),
\eeq
where we have replaced $\beta$ by $j_3$, cf.\ eqn.~\rf{betaj}.
This is precisely the result expected for the circular, elliptic three
spin
string, cf.\ eqn.~\rf{perturbative}. It would of course be interesting to
reproduce the equations~\rf{epsilon} and~\rf{energy} from an exact solution
of eqn.~\rf{intfinal}.

\section{Conclusion}
The continuum $SU(3)$ spin chain sigma model in principle contains all
information about the ${\cal O}(\lambda')$ classical energy of strings
with three angular momenta $(J_1,J_2,J_3)$ on $S^5$ in the limit 
$L=J_1+J_2+J_3\rightarrow \infty$, $\lambda'=\frac{\lambda}{L^2}$ fixed.
Its most general equations of motion are, however, rather involved,
cf.\ \cite{Hernandez:2004uw,Stefanski:2004cw}. It is therefore of interest
to put forward possible simplifying ans\"{a}tze which lead to non trivial
solutions. Previously, it was shown how to recover from the spin chain
sigma model the simple rational three spin string of~\cite{Arutyunov:2003za}.
In the present paper we have presented an ansatz which 
leads to the circular, elliptic three spin string 
of~\cite{Arutyunov:2003uj,Arutyunov:2004,Kristjansen:2004ei}. The most
generic three spin string solutions are parametrized in terms of hyper-elliptic
integrals. It would be interesting to understand how these solutions are encoded
in the spin chain sigma model. Furthermore, it might be that the continuum
spin chain sigma model could reveal solutions overlooked in the string
theory analysis so far.

In the language of the discrete $SU(3)$ spin chain a given three spin
string solution is characterized by a certain Bethe root configuration. 
For the
circular, elliptic three spin string with angular momentum assignment
$(J_1,J_2,J_3)=((1-\alpha)L,(\alpha-\beta)L,\beta L)$ it follows from the
analysis of~\cite{Kristjansen:2004ei} that 
the Bethe root configuration  has to be
of a different type for $\beta<\beta_c(\alpha)$ and
$\beta>\beta_c(\alpha)$ where $\beta=\beta_c(\alpha)$ denotes a line
of critical points in parameter space.
In~\cite{Kristjansen:2004ei} the appropriate Bethe root configuration for
$\beta>\beta_c(\alpha)$ was identified. We propose that the imaginary
root configuration of section~\ref{imaginary} constitutes the
appropriate Bethe root configuration for $\beta<\beta_c(\alpha)$.
Clearly the expression~\rf{gammapert} for the one loop 
anomalous dimension 
as a function
of the spins supports this proposal. In particular, we thus expect
that the imaginary root solution should cease to exist
for $\beta\rightarrow (\beta_c(\alpha))_-$. Certainly, it would be
interesting
to understand the mechanism behind this phenomenon in the spirit of
the understanding of the singular limit
$\beta\rightarrow (\beta_c(\alpha))_+$~\cite{Kristjansen:2004ei}. 
Likewise it would
be interesting to determine the exact location of the critical line.
This would require an  exact solution of the integral
equation~\rf{intfinal}
or of the corresponding integral equation of~\cite{Kristjansen:2004ei}.
We note in passing that neither for the rational three spin string,
nor for the hyper-elliptic one the relevant Bethe root configuration 
is known. 

A recently initiated line of investigation, relying on the observation
that the $SU(3)$ sub-sector may be considered as closed 
in the thermodynamical limit,
is the generalization of the $SU(3)$ spin chain picture to include higher
gauge theory loop orders ~\cite{Minahan:2004ds}. 
A spin chain description going beyond
one loop order was proposed  for the $SU(2)$
sub-sector in~\cite{Serban:2004jf}.  
The corresponding Bethe ansatz implied that inclusion
of higher loop orders required only a rather simple 
modification of the one loop integral equation. In~\cite{Minahan:2004ds}
it was assumed that inclusion of higher 
loop corrections in the $SU(3)$ sub-sector 
lead to a similar modification of the one loop Bethe equations
and the evaluation of higher loop corrections was
carried out for the gauge theory dual of
a circular three spin string with angular 
momentum assignment $(J,J',J')$, $J'<J$.
An exact solution of either of the earlier mentioned integral equations
would allow an extension of the analysis to the case of the more general
circular, elliptic three spin string. 
The study of higher loop corrections has 
so far revealed
a disagreement
between semi-classical string analysis and perturbative gauge theory
at three loop order 
for all cases treated, i.e. 
for folded 
and circular two spin strings~\cite{Serban:2004jf},
a certain class of so-called pulsating strings 
as well as for the above mentioned special
three spin string~\cite{Minahan:2004ds}\footnote{In the case of the two
latter types of strings the disagreement vanishes for a special
choice of parameters which in the case of the three spin string corresponds
to $J=J'$.}. 
A possible explanation
for this discrepancy was proposed in~\cite{Serban:2004jf} and
elaborated in~\cite{Beisert:2004hm}.
Whereas the analysis of the circular, elliptic
three spin string is not expected to change the picture as regards
the presence of the
discrepancy it will provide additional data that might help in 
ultimately resolving it.

\vspace*{0.5cm}

\vspace*{1.0cm}
\noindent
{\bf Acknowledgements}

We are grateful to
Gleb Arutyunov for sharing
with us the unpublished notes~\cite{Arutyunov:2004}.
C.\ K.\ furthermore acknowledges the support of the 
EU network on ``Discrete Random Geometry'', grant HPRN-CT-1999-00161.


\begin{thebibliography}{99}


\bibitem{Frolov:2003qc}
S.~Frolov and A.~A.~Tseytlin,
Nucl.\ Phys.\ B {\bf 668} (2003) 77,
hep-th/0304255.

\bibitem{Frolov:2003tu}
S.~Frolov and A.~A.~Tseytlin,
JHEP {\bf 0307} (2003) 016, hep-th/0306130.

\bibitem{Maldacena:1997re}
J.~M.~Maldacena,
Adv.\ Theor.\ Math.\ Phys.\  {\bf 2} (1998) 231
[Int.\ J.\ Theor.\ Phys.\  {\bf 38} (1999) 1113], hep-th/9711200;
S.~S.~Gubser, I.~R.~Klebanov and A.~M.~Polyakov,
Phys.\ Lett.\ B {\bf 428} (1998) 105, hep-th/9802109;
E.~Witten,
Adv.\ Theor.\ Math.\ Phys.\  {\bf 2} (1998) 253, hep-th/9802150.

\bibitem{Berenstein:2002jq}
D.~Berenstein, J.~M.~Maldacena and H.~Nastase,
JHEP {\bf 0204} (2002) 013, hep-th/0202021.

\bibitem{Kristjansen:2002bb}
C.~Kristjansen, J.~Plefka, G.~W.~Semenoff and M.~Staudacher,
Nucl.\ Phys.\ B {\bf 643} (2002) 3, hep-th/0205033;
%
N.~R.~Constable, D.~Z.~Freedman, M.~Headrick, S.~Minwalla, 
L.~Motl, A.~Postnikov and W.~Skiba,
JHEP {\bf 0207} (2002) 017, hep-th/0205089;
%
N.~Beisert, C.~Kristjansen, J.~Plefka, G.~W.~Semenoff and M.~Staudacher,
Nucl.\ Phys.\ B {\bf 650} (2003) 125, hep-th/0208178;
%
N.~R.~Constable, D.~Z.~Freedman, M.~Headrick and S.~Minwalla,
JHEP {\bf 0210} (2002) 068, hep-th/0209002;
%

\bibitem{Beisert:2002ff}
N.~Beisert, C.~Kristjansen, J.~Plefka and M.~Staudacher,
Phys.\ Lett.\ B {\bf 558} (2003) 229, hep-th/0212269.

\bibitem{Beisert:2003tq}
N.~Beisert, C.~Kristjansen and M.~Staudacher,
Nucl.\ Phys.\ B {\bf 664} (2003) 131, hep-th/0303060.

\bibitem{Minahan:2002ve}
J.~A.~Minahan and K.~Zarembo,
JHEP {\bf 0303}, 013 (2003),
hep-th/0212208.
   
\bibitem{Beisert:2003jj}
N.~Beisert,
Nucl.\ Phys.\ B {\bf 676} (2004) 3,
hep-th/0307015.

\bibitem{Beisert:2003yb}
N.~Beisert and M.~Staudacher,
Nucl.\ Phys.\ B {\bf 670} (2003) 439,
hep-th/0307042.

\bibitem{Lipatov:1993yb}
L.~N.~Lipatov,
JETP Lett.\  {\bf 59} (1994) 596, hep-th/9311037;
L.~D.~Faddeev and G.~P.~Korchemsky,
Phys.\ Lett.\ B {\bf 342} (1995) 311, hep-th/9404173;
V.~M.~Braun, S.~E.~Derkachov and A.~N.~Manashov,
Phys.\ Rev.\ Lett.\  {\bf 81} (1998) 2020, hep-ph/9805225;
A.~V.~Belitsky,
Phys.\ Lett.\ B {\bf 453} (1999) 59, hep-ph/9902361;
V.~M.~Braun, S.~E.~Derkachov, G.~P.~Korchemsky and A.~N.~Manashov,
Nucl.\ Phys.\ B {\bf 553} (1999) 355, hep-ph/9902375;
A.~V.~Belitsky,
Nucl.\ Phys.\ B {\bf 558} (1999) 259, hep-ph/9903512;
Nucl.\ Phys.\ B {\bf 574} (2000) 407, hep-ph/9907420;
S.~E.~Derkachov, G.~P.~Korchemsky and A.~N.~Manashov,
Nucl.\ Phys.\ B {\bf 566} (2000) 203, hep-ph/9909539.

\bibitem{Ferretti:2004ba}
G.~Ferretti, R.~Heise and K.~Zarembo,
hep-th/0404187.
\bibitem{Bethe:1931hc}
H.~Bethe,
Z.\ Phys.\  {\bf 71} (1931) 205.


\bibitem{Beisert:2003xu}
N.~Beisert, J.~A.~Minahan, M.~Staudacher and K.~Zarembo,
JHEP {\bf 0309} (2003) 010,
hep-th/0306139;
N.~Beisert, S.~Frolov, M.~Staudacher and A.~A.~Tseytlin,
JHEP {\bf 0310} (2003) 037,
hep-th/0308117.

\bibitem{Frolov:2003xy}
S.~Frolov and A.~A.~Tseytlin,
Phys.\ Lett.\ B {\bf 570} (2003) 96,
hep-th/0306143.

\bibitem{Serban:2004jf}
D.~Serban and M.~Staudacher,
JHEP {\bf 0406} (2004) 001, hep-th/0401057.

\bibitem{Beisert:2004hm}
N.~Beisert, V.~Dippel and M.~Staudacher,
hep-th/0405001.

\bibitem{Kazakov:2004qf}
V.~A.~Kazakov, A.~Marshakov, J.~A.~Minahan and K.~Zarembo,
JHEP {\bf 0405} (2004) 024, hep-th/0402207;
A.~Marshakov,
hep-th/0406056.


\bibitem{Kruczenski:2003gt}
M.~Kruczenski,
hep-th/0311203.


\bibitem{Kruczenski:2004kw}
M.~Kruczenski, A.~V.~Ryzhov and A.~A.~Tseytlin,
hep-th/0403120.

\bibitem{Beisert:2003ys}
N.~Beisert, Nucl.\ Phys.\ B {\bf 682} (2004) 487,
hep-th/0310252.

\bibitem{Minahan:2004ds}
J.~A.~Minahan,
hep-th/0405243.

\bibitem{Engquist:2003rn}
J.~Engquist, J.~A.~Minahan and K.~Zarembo, JHEP {\bf 0311} (2003) 063,
hep-th/0310188.

\bibitem{Freyhult:2004iq}
L.~Freyhult,
JHEP {\bf 0406} (2004) 010,
hep-th/0405167.

\bibitem{Arutyunov:2003uj}
G.~Arutyunov, S.~Frolov, J.~Russo and A.~A.~Tseytlin,
Nucl.\ Phys.\ B {\bf 671} (2003) 3,
hep-th/0307191.

\bibitem{Arutyunov:2003za}
G.~Arutyunov, J.~Russo and A.~A.~Tseytlin,
Phys.\ Rev.\ D {\bf 69} (2004) 086009,
hep-th/0311004.


\bibitem{Kristjansen:2004ei}
C.~Kristjansen,
Phys.\ Lett.\ B {\bf 586} (2004) 106,
hep-th/0402033.


\bibitem{Hernandez:2004uw}
R.~Hernandez and E.~Lopez,
JHEP {\bf 0404} (2004) 052
hep-th/0403139.

\bibitem{Stefanski:2004cw}
B.~J.~Stefanski and A.~A.~Tseytlin,
JHEP {\bf 0405} (2004) 042, hep-th/0404133.

\bibitem{Arutyunov:2004}
G.\ Arutyunov, unpublished notes.


\bibitem{Kostov:fy}
I.~K.~Kostov,
Mod.\ Phys.\ Lett.\ A {\bf 4} (1989) 217.

\bibitem{Eynard:1995nv}
B.~Eynard and C.~Kristjansen,
Nucl.\ Phys.\ B {\bf 455}, 577 (1995),
hep-th/9506193;
%
Nucl.\ Phys.\ B {\bf 466} (1996) 463,
hep-th/9512052;
B.~Durhuus and C.~Kristjansen,
Nucl.\ Phys.\ B {\bf 483} (1997) 535, hep-th/9609008.

\bibitem{Chekhov:1996xy}
L.~Chekhov and C.~Kristjansen,
Nucl.\ Phys.\ B {\bf 479} (1996) 683,
hep-th/9605013.

\bibitem{Kostov:pn}
I.~K.~Kostov and M.~Staudacher,
Nucl.\ Phys.\ B {\bf 384}, 459 (1992),
hep-th/9203030.

\bibitem{Eynard:1992cn}
B.~Eynard and J.~Zinn-Justin,
Nucl.\ Phys.\ B {\bf 386} (1992) 558,
hep-th/9204082.






\end{thebibliography}
\end{document}